\documentclass[12pt,aps,onecolumn]{revtex4-2}
\usepackage[utf8]{inputenc}
\usepackage[english]{babel}
\usepackage{amsmath}
\usepackage{amsfonts}
\usepackage{amssymb}
\usepackage{dcolumn}
\usepackage{color}
\usepackage{graphicx}				% Use pdf, png, jpg, or eps§ with pdflatex; use eps in DVI mode
\usepackage{amssymb}

\newcommand{\podd}{$\cal P$-odd~}
\newcommand{\ptodd}{$\cal P,T$-odd~}

\newcommand{\ka}{k_{\cal A}~}

\begin{document}
\title{\emph{Ab initio} study and assignment of electronic states in molecular RaCl}
\author{T. A.\ Isaev}
\email{isaev$_$ta@pnpi.nrcki.ru}
\affiliation{National Research Center ``Kurchatov Institute'' - Petersburg Nuclear Physics Institute,
             Orlova Roscha. 1, 188300 Gatchina, Russia}
\author{A. V.\ Zaitsevskii}
\email{zaitsevskii$_$av@pnpi.nrcki.ru}
\affiliation{National Research Center ``Kurchatov Institute'' - Petersburg Nuclear Physics Institute,
             Orlova Roscha. 1, 188300 Gatchina, Russia}
\affiliation{Department of Chemistry, M. Lomonosov Moscow State University, Vorob'evy gory 1/3, Moscow 119991, Russia}
\author{A.\ Oleynichenko}
\affiliation{National Research Center ``Kurchatov Institute'' - Petersburg Nuclear Physics Institute,
             Orlova Roscha. 1, 188300 Gatchina, Russia}
\affiliation{Department of Chemistry, M. Lomonosov Moscow State University, Vorob'evy gory 1/3, Moscow 119991, Russia}
\author{E.\ Eliav}
\affiliation{School of Chemistry, Tel Aviv University, 69978 Tel Aviv, Israel}
\author{A. A.\ Breier}
\affiliation{Laboratory for Astrophysics, Institute of Physics, University of Kassel, 34132 Kassel, Bermany}
\author{T. F.\ Giesen}
\affiliation{Laboratory for Astrophysics, Institute of Physics, University of Kassel, 34132 Kassel, Bermany}
\author{R. F.\ Garcia Ruiz}
%TI - according to Ronald's comments
%\affiliation{CERN, Geneva, Switzerland}
\affiliation{Massachusetts Institute of Technology, Cambridge, MA, USA}
\author{R. Berger}
\affiliation{Fachbereich Chemie, Philipps-Universit\"{a}t Marburg,
  Hans-Meerwein-Strasse 4, 35032 Marburg, Germany}

\date{\today}            
\begin{abstract}
%Abstract version alterred by RB
Radium compounds have attracted recently considerable attention due to both development of experimental techniques
for high-precision laser spectroscopy of molecules with short-lived nuclei and amenability of certain radium compounds for
direct cooling with lasers. Currently, radium monofluoride (RaF) is one of the most studied molecules among the radium compounds,
both theoretically and recently also experimentally. Complementary studies of further diatomic radium derivatives are highly desired to
assess the influence of chemical substitution on diverse molecular parameters,
especially on those connected with laser cooling, such as vibronic transition probabilities, and those related to violations of fundamental
symmetries. In this article
high-precision \emph{ab initio} studies of electronic and vibronic levels of diatomic radium monochloride (RaCl) are presented.
Recently developed approaches for treating electronic correlation with Fock-space
coupled cluster methods are applied for this purpose.
Theoretical results are compared to an early experimental investigation by Lagerqvist and used to partially reassign the
experimentally observed transitions and molecular electronic levels of RaCl.
Effective constants of $\mathcal{P}$-odd hyperfine interaction $W_{\rm{a}}$ and $\mathcal{P,T}$-odd
scalar-pseudoscalar nucleus-electron interaction $W_{\rm{s}}$ in the ground electronic state of RaCl are estimated within the
framework of a quasirelativistic Zeroth-Order Regular Approximation approach and compared to parameters in RaF and RaOH.
%end RB
\end{abstract}
\maketitle

\section{Introduction}

Recently, a new 
%TI later ``laser-spectroscopic'' is used
%spectroscopic 
%end TI
technique has been developed that allows 
to perform high-precision laser-spectroscopic studies of  
molecules with short-lived nuclei \cite{GarciaRuiz:2020}.
The first spectroscopic information on  radium monofluoride (RaF ) molecule was obtained
with this method, including that of isotopologues
with lifetime as short as a few days. These studies have opened an avenue for both fundamental and practical 
investigations with radioactive compounds on numerous radioactive-ion-beam facilities around the world. 
The experimental scheme heavily relies on
information from electronic structure modeling for the corresponding compounds, and in this connection 
calculations of the electronic structure of RaX systems, where X is a 
group 16 or 17 element of the Periodic Table attract considerable 
attention: In such compounds one could expect considerable enhancement
of $\mathcal{P}$- and $\mathcal{T}$-odd effects, including those connected with nuclear spin (see \cite{Kozlov:95, Kudashov:12, Isaev:10, Isaev:13} and references therein) as well as the existence of highly closed optical cycling loops,
which make such molecules amenable for direct laser cooling \cite{Isaev:10}. Thus, radium compounds are 
especially attractive for experimental search for signatures of new physics, 
including $\mathcal{P}$- and $\mathcal{T}$-odd effects.   
The radium monochloride molecule (RaCl) is of
special interest as it allows to observe the influence of halogen substitution in RaX compounds
on various molecular parameters, particularly on those connected with laser cooling, 
such as Franck--Condon (FC) factors and oscillator strengths, as well as  on $\mathcal{P}$- and $\mathcal{T}$-odd effects.
The precursor material radium dichloride is more readily available on the bulk level, which is important for future
off-line spectroscopic studies. Moreover, early experimental data on optical spectra of RaCl 
%Ronald's comment
containing the long-lived isotope $^{226}$Ra
exist, which date back to 1953 \cite{Lagerqvist:53} and motivate the present 
%TI
study (during the course of our work on the present manuscript, we became aware of another theoretical study on RaCl \cite{Osika:2020} triggered by discussions of T. Isaev with the authors of that manuscript).
%end TI
As RaF, the RaCl molecule is also an excellent object to test modern methods of relativistic electronic structure calculations.
To study ground and excited electronic states of RaCl with high accuracy, we employ 
the Fock-space relativistic coupled cluster method (\cite{Visscher:01} and references therein). 
Transition probabilities and lifetimes of the excited states are computed within
the finite-field technique~\cite{Zaitsevskii:98,Zaitsevskii:18}, 
successfully applied earlier to a number of heavy-atom 
compounds~\cite{Isaev:17,Medvedev:17}. Additionally, we estimate effective constants of 
$\mathcal{P}$-odd hyperfine interaction $W_{\rm{a}}$ and $\mathcal{P,T}$-odd scalar electron-nucleus
interaction $W_{\rm{s}}$ (see \cite{Kozlov:95} for definition of the constants) of the ground electronic 
$^2\Sigma_{1/2}$ state of RaCl
in the framework of Zeroth-Order-Regular-Approximation combined with Generalized Hartree-Fock or Generalized Kohn-Sham approach
 (ZORA/GHF and ZORA/GKS respectively).  

\section{Computational details}

\paragraph{Excited state calculations.} 
Accurate \emph{ab initio} evaluation of transition energies in RaCl implies an adequate description
of relativistic effects, including those beyond the conventional Dirac--Coulomb model. 
%az
%The importance 
%of incorporating the Breit interaction follows from its effect on the lowest ($7s-6d$) excitation energies in
%the atomic Ra${}^+$ ion~\cite{Eliav:96}. 
%
%(\emph{I am not quite convinced here, because Eliav et al. write that the Breit contribution to energies
%is small, only fine structure splitting is improved. If you look at the numbers in the paper, Breit
%introduces a change of less than 100/cm. But then in section III of the present manuscript it is
%written that "the slight (210-220/cm) overestimation of $7s-6d$ excitation energies in the free Ra+
%atomic ion within the present electronic structure model." That "slight" deviation, however, is
%much larger than the Breit contribution we are talking about.})
%
%az We have several sources of errors, finite basises, neglected cluster amplitudes etc, and having a significant systematic error with clear origin, 
%even if it not a dominating one, we have to correct it; the remainder error 210-220 cm-1 seems better than 300+ cm-1. 
%Maybe it is better to replace the text above by somewhat like
The non-negligible effect of non-Coulombian (Breit) electron-electron interactions on the lowest ($7s-6d$) excitation energies in the atomic Ra${}^+$ ion had been demonstrated in~\cite{Eliav:96}.
%end az

Many modern relativistic all-electron codes for electron correlation calculations in molecules, however,
do not account for the Breit interaction.
Furthermore, the secondary relativistic expansion and destabilization of
high-angular-momentum one-electron states of Ra leads to the necessity to correlate, in addition to the valence 
and eight subvalence ($6sp$) electrons, also the electrons associated with the $5d$-shell. Additional
difficulties stem from a very slow convergence of $6d$-state correlation energies for Ra$^{+}$ with the extension
of orbital basis sets; preliminary estimates have shown that the basis set should preferably include
functions with orbital angular momentum quantum numbers $l$ of up to 6 ($i$-functions).  
Taking into account above-mentioned challenges we are using in the present work the relativistic electronic structure model defined by
accurate two-component pseudopotentials of ``small'' atomic cores derived from the valence-shell solutions  
of the atomic Dirac--Fock--Breit equations. Implicitly incorporating the bulk of Breit interaction effects
and leaving the electrons of outer core shells for explicit correlation treatment, this model also
offers the possibility of using sufficiently large and flexible contracted Gaussian basis sets. 
 
In the present study, the pseudopotentials~\cite{Mosyagin:10a,ourGRECP} replaced the shells with the principal 
quantum number $n\le 4$ for Ra and the $1s$-shell for Cl.  
The basis set for Ra comprised a set of primitive Gaussians $(10s\,9p\,9d\,7f)$, loosely based on the diffuse part of 
the primitive function set from Ref.~\cite{Widmark:04}, and the contracted Gaussians 
$(7g\,6h\,5i)/[4g\,3h\,2i] $ obtained as averaged atomic natural orbitals in scalar-relativistic calculations
on low-lying electronic states of Ra and Ra$^+$. The $(10s\,12p\,4d\,3f\,2g)/[6s\,7p\,4d\,3f\,2g]$ chlorine basis was a straightforward adaptation of the aug-cc-pVQZ set~\cite{Woon:93} to the use with the pseudopotential, extended by
the additional single set of contracted $p$-functions ($2p_{3/2}-2p_{1/2}$) to improve the description of spin-orbit splittings.
% RB
%Exponent coefficients and linear expansion coefficients for the contracted basis sets are given in the appendix.
%\emph{(I doubt that anyone will be able to reproduce these settings, if we do not give all data in the appendix/supplement)}
%az
%obviously. In supplement (“Supplementary material” in JQSRT, e.g.) rather than in Appendix, otherwise it's too heavy; I would prefer 
Exponential parameters and linear expansion coefficients for the contracted basis sets can be found on the website ~\cite{ourGRECP} and are given in the Supplementary materials.
%TI we are preparing Supplementary with AZ, for now my raw data on ZORA calculations are in $proj/diatomic_pv/RaCl
%end az
%RB end

Correlation calculations were performed using the FS RCC method~\cite{Visscher:01}. 
Molecular (pseudo) spinors were generated by solving the SCF 
equations for the closed-shell ground state of ${\rm RaCl}^{+}$ which was used as Fermi vacuum state.
The ground and low-lying excited states of neutral RaCl thus corresponded to the $0h1p$
sector of the Fock space. The cluster operator
expansion comprised only single and double excitations. 
27 electrons (20 electrons of Ra, including those of subvalence $5d\,6sp$ shells, and 7 valence electrons of Cl) were correlated whereas the spinors corresponding to the $5sp$ shells Ra and $2sp$ shells Cl 
were frozen after the Hartree--Fock step. 
In most calculations the model space in the target $0h1p$ sector was defined by 9 Kramers pairs of 
``active'' spinors.

To reduce systematic internuclear-distance-dependent
errors in the energy values arising from the neglect of higher cluster amplitudes and the basis set superposition (BSSE)~\cite{Pazyuk:15} the potential energy curves for the excited states were constructed by adding the FS RCC vertical excitation energies calculated as functions of the internuclear distance, to the ground state potential recomputed with perturbative account for the contribution from triple excitations 
(RCCSD(T) scheme) and 
% RB
%counterpoise BSSE corrections.
with counterpoise corrections of the BSSE. 
% RB end ... otherwise it is not clear, if also BSSE was only corrected perturbatively.
The calculations were performed 
% RB
% for the internuclear separation $R$ in the range  $4.6\le R \le 6.5$ a.u.,
for internuclear separations $R$ in the range 
$4.6~a_0\le R \le 6.5~a_0$, 
% RB end
where the validity of the chosen electronic structure model (restriction to the single Fock space sector $0h1p$) for the low-lying states $(1-5)\,1/2$, $(1-3)\,3/2$, and $(1)\,5/2$ was not in doubt; at the same time the 
%RB 
% obtained  fragements of potential curves were
chosen coordinate range for the potential energy curves was 
%RB end
sufficiently large for highly accurate numerical evaluation of wavefunctions and 
energies of 4$-$5 lowest vibrational levels for each of these states.

In order to get insight into the impact of spin-dependent relativistic effects on the composition of the excited
state wavefunctions, we computed the scalar relativistic counterparts of the states under study 
\emph{via} switching off the spin-orbit parts of the pseudopotential
and projected these states onto the model-space parts of fully relativistic states (cf.~\cite{Zaitsevskii:17}).

%RB
%Transition dipole matrix elements 
Matrix elements of the electric transition dipole moment
% RB end
between the low-lying electronic states were evaluated
using the finite-field technique~\cite{Zaitsevskii:98,Isaev:17,Zaitsevskii:18}.
In contrast to total energies, the finite-field transition dipole estimates
are significantly improved upon extension of the model space~\cite{Zaitsevskii:18}, so here
we used a larger model space corresponding to 19 Kramers pairs of active spinors. Numerical instabilities
which appeared for this extended model space were eliminated by applying small ``dynamic'' shifts of FS RCC energy denominators~\cite{Zaitsevskii:17} in the target sector 
(shift amplitudes $-0.05~E_\mathrm{h}$ and $-0.10~E_\mathrm{h}$ for single and double excitations, respectively, and the attenuation parameter
$m=3$, see Eq.~(4) in Ref.~\cite{Zaitsevskii:17}).

The Kramers-restricted SCF and single-reference RCCSD(T) ground state calculations 
were performed with the DIRAC program suite~\cite{DIRAC:17}. 
%AZ
%For FS RCC calculations we
%employed the EXP-T program recently developed by some of 
%the authors of the present work (A.O., A.Z., and E.E.)~\cite{EXPT}. 
% The EXP-T code design employs the fact that CC amplitude equations can be represented by tensor contractions, implemented as matrix multiplications (for details, see 
%\cite{Crawford:07,Bartlett:09,Solomonik:14}); such an approach results in easy extensibility of the code. Molecular symmetry is handled via the direct-product decomposition %(DPD) approach \cite{Stanton:91,Matthews:19} extended to the case of relativistic double groups \cite{Visscher:96a,Shee:16}. EXP-T is parallelized for both multi-core CPUs %(via the OpenMP technology \cite{OpenMP}) and NVIDIA GPUs (via the CUDA technology \cite{CUDA,CUDAweb}). The recently proposed energy denominator shift 
%technique \cite{Zaitsevskii:17,Zaitsevskii:18a} is also implemented being the unique feature of \mbox{EXP-T}. The codes for finite-field calculations of transition dipole %moments and intensities \cite{Zaitsevskii:18} and Pad\'e extrapolations of series of effective Hamiltonians \cite{Zaitsevskii:18a} are also included in the EXP-T package.
For FS RCC calculations we
employed the EXP-T program recently developed by some of 
the authors of the present work (A.O., A.Z., and E.E.)~\cite{EXPT,EXPT:20}. 
%end AZ
Transformed molecular integrals were calculated with the help of the DIRAC
18 program suite and then imported to EXP-T.

Vibrational wavefunctions and energy levels as well as FC factors were evaluated with the help of the program 
VIBROT~\cite{Sundholm}.

\paragraph{Calculations of $W_{\rm{a}}$ and $W_{\rm{s}}$ constants.}
The \podd nuclear spin-dependent and  \ptodd scalar electron-nucleus contributions for the
given nucleus $A$ (in our case the Ra nucleus) in paramagnetic diatomic molecules
are described by the following terms of the
effective molecular Hamiltonian \cite{Flambaum:85, Kozlov:95}:
\begin{eqnarray}
 W_\mathrm{a} \ka [\vec{\bm{\lambda}}{\times}\vec{\bf S}^\mathrm{eff}]\cdot \vec{\bf I} +
W_\mathrm{s} k_{\mathrm{\cal SP}}~ \vec{\bm{\lambda}} {\cdot} \vec{\bf S}^\mathrm{eff},
\label{heff}
\end{eqnarray}
where $\vec{\bm{\lambda}}$ is the unit vector directed from Ra to Cl,
$\vec{\bf S}^\mathrm{eff}$ is the effective electronic spin and $W_\mathrm{a}$ and $W_\mathrm{s}$
are parameters, which have to be deduced from electronic structure calculations. The physical meaning of the
$W_\mathrm{a}$ parameter is \podd hyperfine coupling constant \cite{flambaum:1984,Isaev:2014},
while $W_\mathrm{s}$ can be related to the parameters of \ptodd electron paramagnetic resonance. 
The computation scheme of \podd and \ptodd parameters is analogous
to that used in \cite{Isaev:12, Isaev:2014}.
% RB
%TI removed Isaev:16
%(\emph{Reference \cite{Isaev:16} does not fit here}).
%
We used for Ra basis set of uncontracted Gaussian with
the exponent coefficients composed in the even-tempered manner as in \cite{Isaev:2014}.
% RB
%TI removed Isaev:16
%(\emph{Reference \cite{Isaev:16} does not fit here}).
%
For Cl basis set of triple-zeta quality augmented with polarisation $d$-functions
% RB
%(\emph{TZz is not contained in the TM basis set library})
%TI Good point, it is in tm2c basis set library. In any case Cl basis set will be given in Supplementary, 
%but in text it has to be changed
%from TURBOMOLE basis set library (TZz) is used.
from {\sc tm2c} basis set library (TZz) is used (see Supplementary for explicit basis set exponents).
%
%----------------------------------------------------------------------------------
The nuclear density was modelled by a spherical Gaussian distribution $\rho
(R)=\rho_0 e^{-\frac{3}{2\xi}R^2}$, where $\xi$ is the mean root square
radius of the corresponding nucleus computed according to the empirical
formula $\xi=(0.836 A^{1/3}+0.57)~\mathrm{fm} = (1.5798
A^{1/3}+1.077)~10^{-5}a_0$, where $A$ is taken equal to be 226 for Ra.
A modified version
\cite{Berger:2005,Berger:2005a,Nahrwold:09,Isaev:12} of the program
package {\sc TURBOMOLE} \cite{Alrichs:89} was used for the 
%RB
%GHF (Hartree--Fock or Kohn--Sham) calculations.
generalised Hartree--Fock (GHF) or generalised Kohn--Sham (GKS) calculations.
%RB end
The value of \podd and \ptodd parameters are calculated at equilibrium internuclear distance obtained
in 
%RB
% geometry optimisation
structure energy-minimisations 
%RB end
performed  by {\sc TURBOMOLE} using 68-electron Stuttgardt effective core potential
on Ra with TZVP basis set on Ra [9s,9p,5d,3f] and TZP basis set on Cl [9s,5p,1d]. The optimised $R_\mathrm{e}$ value is
$5.361~a_0$ which is quite close to the $R_\mathrm{e}$ from RCCSD(T) calculations ($5.271~a_0$).
%RB
%TI given
%(\emph{I would suggest to give the CCSD(T) value here then as well, because later it is just
%reported in Angstrom, not in units of the Bohr radius})
% RB end
The parameter $|W_\mathrm{a}|$ was calculated according to the Eq. (3) from \cite{Isaev:12},
and the parameter $W_\mathrm{s}$ according to the Eqs. (2) -- (4) from \cite{Isaev:13}
(we note that due to a typo the $\rho_A(\vec{r})$ member was omitted in expression (2) in \cite{Isaev:13}).
We have neglected the contributions from the relatively light nuclei of Cl to the considered molecular
\podd and  \ptodd properties,
due to the strong dependence of these effects on the nuclear charge (see e.g. \cite{Isaev:12} and
references therein).
We note that in the direct application of the
present complex GHF/GKS approach only the absolute value of $W_\mathrm{a}$
is immediately accessible, while for $W_\mathrm{s}$ the sign is determined directly in the calculations.

\section{Results and discussion}
\begin{figure}
  \centering
    \includegraphics[width=1.0\textwidth]{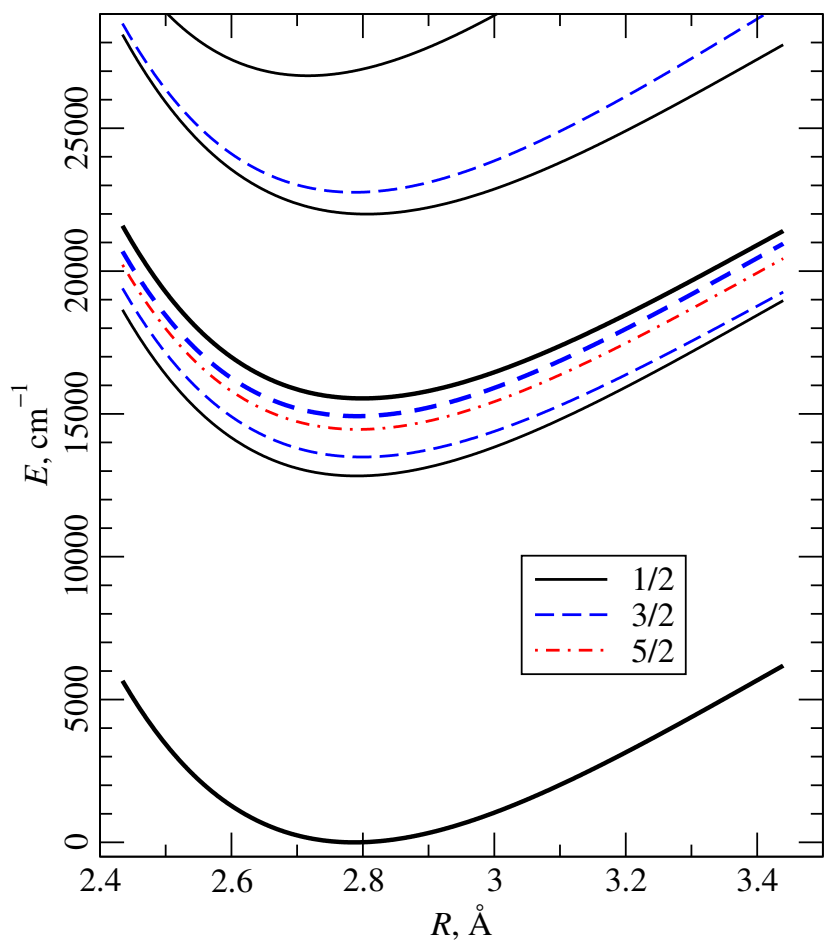}
  \caption{Calculated potential energy functions for the low-lying electronic states of RaCl. 
  Bold lines correspond to the supposed counterparts of the experimentally observed states $X$, $C$, and $C'$. }
\label{curves}
\end{figure}

The resulting potential energy functions are presented in Fig.~\ref{curves} and the derived values 
for the main molecular constants of RaCl along with the corresponding spectroscopic data
from Ref.~\cite{Lagerqvist:53} and estimated transition moments to the ground states are listed in 
Table~\ref{spectro}.  
 The low-lying excited states, $(2-3)\,1/2$, $(1-2)\,3/2$, and $(1)\,5/2$, having very similar shapes of interatomic potentials, 
 form a rather tight group which  
 can be naturally associated with $^2D_J$-states of Ra$^+$ perturbed by the interaction
 with the chlorine anion.  
 The parameters of two states with bright transitions from the ground one, namely, the 
 second state with $\Omega=3/2$ and third state with $\Omega=1/2$, are in excellent agreement 
 with those of the experimentally observed states $C$ and $C'$ \cite{Lagerqvist:53}. The small deviations
 of the computed $T_\mathrm{e}$ values from their experimental counterparts (+139 cm$^{-1}$ for and +154 cm$^{-1}$ respectively) 
 correlate with the slight (210--220 cm$^{-1}$) overestimation 
 of $7s-6d$ excitation energies in the free Ra$^+$ atomic ion within the present 
 electronic structure model.  
 
 %ion RaCl+
  The present model provides directly the RCCSD(T) estimate for the RaCl ground-state dissociation energy 
 to the ionic
 fragments, ${\rm Ra}^+\,(7s\,^{2\!}S_{1/2})$ and Cl$^-\,(^1S_0)$, $47\,680 \,{\rm cm}^{-1}$. Combining this
 estimate with the accurate experimental ionization energy of atomic Ra, $42\,573\;{\rm cm}^{-1}$ \cite{NIST_asd:19} 
 and electron
 affinity of chlorine, $29\,139\;{\rm cm}^{-1}$ \cite{Rienstra:02}, one readily obtains the RaCl adiabatic 
 dissociation energy $D_\mathrm{e}=34\,246\;{\rm cm}^{-1}$ (ca. 4.25 eV). The FS RCCD
 calculation on Ra ionization potential (with the vacuum state redefined as $7s^2\;^1S_0$) and chlorine
 electron affinity yields the values $42\, 913\; {\rm cm}^{-1}$ and $29\, 477\; {\rm cm}^{-1}$, respectively,
 which can be used to evaluate the RaCl ($X\,1/2$) dissociation energy to the lowest-energy channel without 
 any reference to empirical data ($D_\mathrm{e}=34\; 244\;{\rm cm}^{-1}$).
 %end ion
 
\begin{table}[h!]
 \caption{\small Molecular constants,  
  squared electric dipole transition moments for transition to the ground state $|d|^2$
  and the composition of the model wavefunctions in terms of their scalar relativistic counterparts at $R_\mathrm{e}(X)$ for low-lying electronic states of RaCl.
  }
\begin{center}
  \begin{tabular}{lcccccc}
             & \raisebox{-2ex}{$T_\mathrm{e}$, cm$^{-1}$ } & \raisebox{-2ex}{$R_\mathrm{e}$, {\AA}}&  \multicolumn{2}{c}{$\omega_\mathrm{e}$, cm$^{-1}$} & \raisebox{-2ex}{$|d|^2$, a.u.}  & \raisebox{-2ex}{Composition} \\ 
             &         &         & $^{226}$Ra$^{35}$Cl&   $\!\! ^{226}$Ra$^{37}$Cl    \\\hline
$ X(1)1/2$     & 0  &  2.789  & 258.5  & 252.5  & & $100\%\,^2\Sigma $\\
%$ X $, exptl \cite{Lagerqvist:53} & 0 & & \multicolumn{2}{c}{256.2$^{*)}$}  \vspace{1.ex}  \\   
%% Added by RB based on data communicated by Alex Breier
%$ X $, exptl, re-analysis 1$\dagger$& 0 & & \multicolumn{2}{c}{256.48(14)$^{*)}$}  \vspace{1.ex}  \\
%$ X $, exptl, re-analysis 2$\dagger$& 0 & & 262.45(16) & & & \\
%
%\\
(2)1/2        & 12827    & 2.792 & 259.0      & 253.0    & 4.62 & $26\%\,^2\Sigma+74\%\,^2\Pi$      \vspace{1.ex}       \\ 
(1)3/2        &  13487   & 2.800 &253.9 &       248.0 &0.21
& $8\%\, ^2\Pi + 92\%\, ^2\Delta$  \vspace{1.ex} \\
(1)5/2       &  14454  &  2.794 & 256.3 & 250.3 & 0 & $100\%\,^2\Delta$  \vspace{1.ex} \\
(2)3/2        & 14921  & 2.793 &  257.2  &  251.2 & 4.03 & $92\% \,^2\Pi+8\% \,^2\Delta$ \\
%$ C $, exptl \cite{Lagerqvist:53} & {}\hspace{1.7ex}14782.1 &  & \multicolumn{2}{c}{253.8$^{*)}$} \vspace{1.ex}    \\
%% Added by RB based on data communicated by Alex Breier
%$ C $, exptl, re-analysis 1$\dagger$ & {}\hspace{1.7ex}14786.57(11) &  & \multicolumn{2}{c}{254.01(14)$^{*)}$} \vspace{1.ex}    \\
%$ C $, exptl, re-analysis 2$\dagger$ & {}\hspace{1.7ex}14786.57(12) &  & 259.98(16) & &    \\
%
(3)1/2        &  15541  &  2.800 & 256.6 & 250.6  & 3.33 &  $74\% \, ^2\Sigma + 26\% \,^2\Pi$  \\
%$ C'$, exptl \cite{Lagerqvist:53} & {}\hspace{1.7ex}15386.5 & & \multicolumn{2}{c}{252.9$^{*)}$}  \vspace{1.ex}   \\
%% Added by RB based on data communicated by Alex Breier
%$ C'$, exptl, re-analysis 1$\dagger$ & {}\hspace{1.7ex}15390.75(10) & & \multicolumn{2}{c}{253.18(12)$^{*)}$}  \vspace{1.ex}   \\
%$ C'$, exptl, re-analysis 2$\dagger$ & {}\hspace{1.7ex}15390.73(11) & & 259.15(14) & &  \\
%
(4)1/2        & 21991  &  2.807 & 258.0 &    251.9 &2.16& $99\% \,^2\Pi$ \vspace{1.ex} \\
(3)3/2        & 22758 & 2.789 &265.6 &  259.4 & 2.43 & $100\% \,^2\Pi$ \vspace{1.ex} \\
(5)1/2        & 26852 & 2.718 &274.9    &   268.3&0.18 & $99\% \,^2\Sigma$ \\
%ion
\multicolumn{7}{c}{ RaCl$^+$ }\\
$\tilde{X}, 0^+$& 42168 & 2.672 & 294.9  &    287.7 & \\
  \end{tabular}
\end{center}
 % $^{*)}$ Unknown isotopologue     \hfill                                             
%% Added by RB based on data communicated by Alex Breier
  %$^{\dagger)}$ Additional parameters from re-analysis 1: $\omega_\mathrm{e}x_\mathrm{e}/\mathrm{cm}^{-1}$ 
    %            for states X, C and C': 0.732(12), 0.731(12), 0.744(11);
     %           Additional parameters from re-analysis 2: $\omega_\mathrm{e}x_\mathrm{e}/\mathrm{cm}^{-1}$
      %          for states X, C and C': 0.734(19), 0.733(19), 0.746(18).
%
  \label{spectro}
\end{table}
%TI
\begin{table}[h!]
 \caption{\small Molecular constants for low-lying electronic states of RaCl according to \cite{Lagerqvist:53}
 and our re-analysis of experimental data.
  }
\begin{center}
  \begin{tabular}{lcccccc}
             & \raisebox{-2ex}{$T_\mathrm{e}$, cm$^{-1}$ } & \raisebox{-2ex}{}&  \multicolumn{2}{c}{$\omega_\mathrm{e}$, cm$^{-1}$} & \raisebox{-2ex}{}  & \raisebox{-2ex}{} \\
                         &         &         & $^{226}$Ra$^{35}$Cl&   $\!\! ^{226}$Ra$^{37}$Cl    \\\hline 
$ X $, exptl \cite{Lagerqvist:53} & 0 & & \multicolumn{2}{c}{256.2$^{*)}$}  \vspace{1.ex}  \\   
% Added by RB based on data communicated by Alex Breier
$ X $, exptl, re-analysis 1$\dagger$& 0 & & \multicolumn{2}{c}{256.48(14)$^{*)}$}  \vspace{1.ex}  \\
$ X $, exptl, re-analysis 2$\dagger$& 0 & & 262.45(16) & & & \\
\hline
$ C $, exptl \cite{Lagerqvist:53} & {}\hspace{1.7ex}14782.1 &  & \multicolumn{2}{c}{253.8$^{*)}$} \vspace{1.ex}    \\
% Added by RB based on data communicated by Alex Breier
$ C $, exptl, re-analysis 1$\dagger$ & {}\hspace{1.7ex}14786.57(11) &  & \multicolumn{2}{c}{254.01(14)$^{*)}$} \vspace{1.ex}    \\
$ C $, exptl, re-analysis 2$\dagger$ & {}\hspace{1.7ex}14786.57(12) &  & 259.98(16) & &    \\
\hline
$ C'$, exptl \cite{Lagerqvist:53} & {}\hspace{1.7ex}15386.5 & & \multicolumn{2}{c}{252.9$^{*)}$}  \vspace{1.ex}   \\
% Added by RB based on data communicated by Alex Breier
$ C'$, exptl, re-analysis 1$\dagger$ & {}\hspace{1.7ex}15390.75(10) & & \multicolumn{2}{c}{253.18(12)$^{*)}$}  \vspace{1.ex}   \\
$ C'$, exptl, re-analysis 2$\dagger$ & {}\hspace{1.7ex}15390.73(11) & & 259.15(14) & &  \\
\hline
%end TI
 \end{tabular}
\end{center}
  $^{*)}$ Unknown isotopologue     \hfill                                             
% Added by RB based on data communicated by Alex Breier
  $^{\dagger)}$ Additional parameters from re-analysis 1: $\omega_\mathrm{e}x_\mathrm{e}/\mathrm{cm}^{-1}$ 
                for states X, C and C': 0.732(12), 0.731(12), 0.744(11);
                Additional parameters from re-analysis 2: $\omega_\mathrm{e}x_\mathrm{e}/\mathrm{cm}^{-1}$
               for states X, C and C': 0.734(19), 0.733(19), 0.746(18).
  \label{spectro-exp}
\end{table}

 Making use of the analogy between RaCl and the monochlorides of lighter group II elements, 
 Lagerqvist had identified the $C,C'$ states with the components
 of the spin--orbit-split $^2\Pi$ manifold \cite{Lagerqvist:53}. 
 %RB
 The splitting of $604.4~\mathrm{cm}^{-1}$ between the two levels observed in experiment 
 would, indeed, be essentially consistent with the expected spin--orbit splitting in the $C$ state 
 of RaCl. But this level is expected --- and also computationally found --- at higher and not lower
 energies than the corresponding level in BaCl: It correlates with the $7p$ level of Ra$^+$,
 and the $np$ level in the alkaline earth monocations is known to drop in the sequence from 
 Be$^+$ to Ba$^+$, but raises again for Ra$^+$.
 %RB end
 %
 %The assignment does not make much sense since the $\Lambda\!-\!S$ coupling scheme, being reasonable
 %for lighter group II monochlorides, is broken down for RaCl. 
 %Indeed, the separations between the
 The $\Lambda\!-\!S$ coupling scheme is also partially broken for RaCl. 
 The separations between the
 %RB end
 scalar relativistic excited $^2\Delta$,$^2\Pi$, and $^2\Sigma$ terms in the vicinity of equilibrium
 are about 140 and 800 $\rm cm^{-1}$, respectively, being much smaller than the spin-orbit splitting 
 of the $^2D$ level in Ra$^+$ (1659   $\rm cm^{-1}$) characterizing the intensity of spin-dependent
 interactions within this group of states. The analysis of the model-space wavefunctions
 confirms the domination of $^2\Pi$-like component in the supposed $C$ state, $(2)\,3/2$, whereas the
 supposed $C'$ state, $(3)\,1/2$, is a $^2\Sigma$ - $^2\Pi$ admixture with a predominance of $^2\Sigma$ 
 (see the last column of Table~\ref{spectro}).  

% Added by RB based on the reanalysis communicated by Alex Breier
 We made an attempt to reanalyse the list of transitions reported by Lagerqvist, assuming that the
 band heads reported as Q1 and Q2 in Lagerqvists work stem from independent electronic states
 rather than the spin-orbit split ${}^2\Pi$ level. The corresponding molecular constants are given
 %TI
 %in table~\ref{spectro} 
 in Table~\ref{spectro-exp} 
 %end TI
 as reanalysis 1. Reanalysis 2 takes additionally into account that ${}^{35}$Cl
 and ${}^{37}$Cl have a natural abundance ratio of approximately 3:1. For the $\Delta\nu=0$ branch,
 one would not expect to observe the difference between isotopic species in a measurements with a resolution of 0.01~\AA~or
 less. But in the $\Delta\nu=-1$ branch, one would expect to distinguish between the two isotopologues
 due to the lower abundance of ${}^{37}$Cl. As only band head positions, but not the intensities were
 reported by Lagerqvist, one can only tentatively assume that some of the transitions reported correspond
 to the ${}^{37}$Cl containing RaCl isotopologue. In reanalysis 2, the first three transition were assumed
 to belong to $^{226}$Ra${}^{37}$Cl, and the following transitions are assumed to be dominated by $^{226}$Ra${}^{35}$Cl, 
 which leads to a relabeling of the vibrational quantum number. The constants resulting for the latter 
 isotopologue are given in 
 %TI
 Table~\ref{spectro-exp}.
 %end TI

 Lagerqvist mentioned already in his early analysis that the Franck--Condon parabola is very narrow 
 for RaCl, which is a prerequisite for laser-coolability.
 In accordance with the experimental finding,
%end
 the potential curves of all states of the group under discussion are nearly parallel to that of the ground state;
 the equilibrium internuclear separation and bond strength remain almost unchanged upon electronic excitation. 
  %RB
 %This points on possibility of (quasi)closeness of optical 
 This indicates the possibility for (quasi)closed optical 
 %RB end
 %pump-dump 
  cycles involving these states. We calculated the sums of  three largest FC-factors  for decay
  from the lowest ($v'=0$) vibrational level of (2)1/2,  (1)3/2,   (2)3/2, and  
 (3)1/2 to 
 %RB
 the
 %EB end
 lowest vibrational levels of the ground state. The results (Table~\ref{sumffc})
 indicate that at least for the lowest excited state $(2)\,1/2$  the effective cooling cycle
  could be build with at most three lasers (one basic laser on $0'-0$ transition and two repumpers on $0'-1$ and $0'-2$). 
   
\begin{table}
\caption{\small Sums of the FC factors for transitions from the lowest
% RB
% ($v'=0$) vibrational level of excited states to several first vibrational levels ($v=n$) of the ground
%electronic state, $\sum_{n}\left<v'=0\vert v=n\right>^2$, for  $^{226}$Ra$^{35}$Cl. 
 ($v'=0$) vibrational level of excited states to several vibrational levels ($v$) of the ground
electronic state, $\sum_{i=0}^n|\left<v'=0\vert v=i\right>|^2$, for  $^{226}$Ra$^{35}$Cl. 
% RB end
The  corresponding values for $^{226}$Ra$^{37}$Cl are nearly the same. } 
 \begin{center}
  \begin{tabular}{lccc}
\\
%TI
% State    &  $V_X''=0$ &  $V_X''=1$ &  $V_X''=2$ \\ \\
Initial state    &  $n=0$ &  $n=1$ &  $n=2$ \\ \hline
%end TI
(2)1/2             & 0.998516 & 0.999995 &1.000000 \\
(1)3/2             & 0.983494 & 0.999911 &1.000000 \\
$C$ (2)3/2         & 0.997818 & 0.999999 &1.000000  \\
$C'$ (3)1/2        & 0.983924 & 0.999968 & 1.000000\\\hline
  \end{tabular}                                                  
  \end{center}
  \label{sumffc}
\end{table}

The closeness of optical cycles involving higher-lying 
excited states can be destroyed by spontaneous radiative transitions 
%RB
and internal conversion
%RB end
to lower excited states. The radiative branching rates derived from FS RCC 
electric transition dipole moments
and energies
(Table~\ref{branching}) show that the relative probabilities of the decay of the (1)3/2, (2)3/2, and (3)1/2 states to lower excited states are rather
small (about $10^{-3}$ for the former and $10^{-4}$ for two latter states).
Moreover, due to the quasidiagonality of FC matrices for deexcitations of all states belonging to the manifold under discussion, the transition to some ``intermediate'' state should hardly break the cycle: 
%RB
%highly probably, 
with high probability
%RB end
the radiative decay of such state will still lead to one of the lowest vibrational levels of the ground state. 
Thus not only the lowest-energy excitation to the $(2)\,1/2$ state, but also the next three electric dipole-allowed excitations
(including those to the experimentally observed  $C$ and $C'$ states) can be considered as prospective candidates
for implementing efficient cooling optical cycles.

\begin{table}
\caption{\small Estimates of relative spontaneous radiative decay branching ratios for several excited states of RaCl. } 
 \begin{center}
  \begin{tabular}{ll<{\hspace{3ex}}l}
\\
Initial state & \multicolumn{1}{c}{Final states} & \multicolumn{1}{c}{Branching ratios} \\\hline
(1)3/2 & $\to\!\! X$ : $\to\!\! (2)1/2$             & 1 :  $1.2\!\cdot\! 10^{-3}$\\
(2)3/2 & $\to\!\! X$ : $\to\!\! (2)1/2$ : $\to\!\! (1)3/2$ : $\to\!\! (1)5/2$   & 1 : $1.3\!\cdot\! 10^{-5}$   : $2.5\!\cdot\! 10^{-5}$ : $2.3\!\cdot\! 10^{-5}$\\
(3)1/2 & $\to\!\! X$ : $\to\!\! (2)1/2$ : $\to\!\! (1)3/2$ : $\to\!\! (2)3/2$   & 1 : $3.9\!\cdot\! 10^{-4}$ : $2.7\!\cdot\! 10^{-4}$ : $1.0\!\cdot\!  10^{-5}$\\\hline
  \end{tabular}                                                  
  \end{center}
  \label{branching}
\end{table}

\begin{figure}
  \centering
    \includegraphics[width=0.7\textwidth]{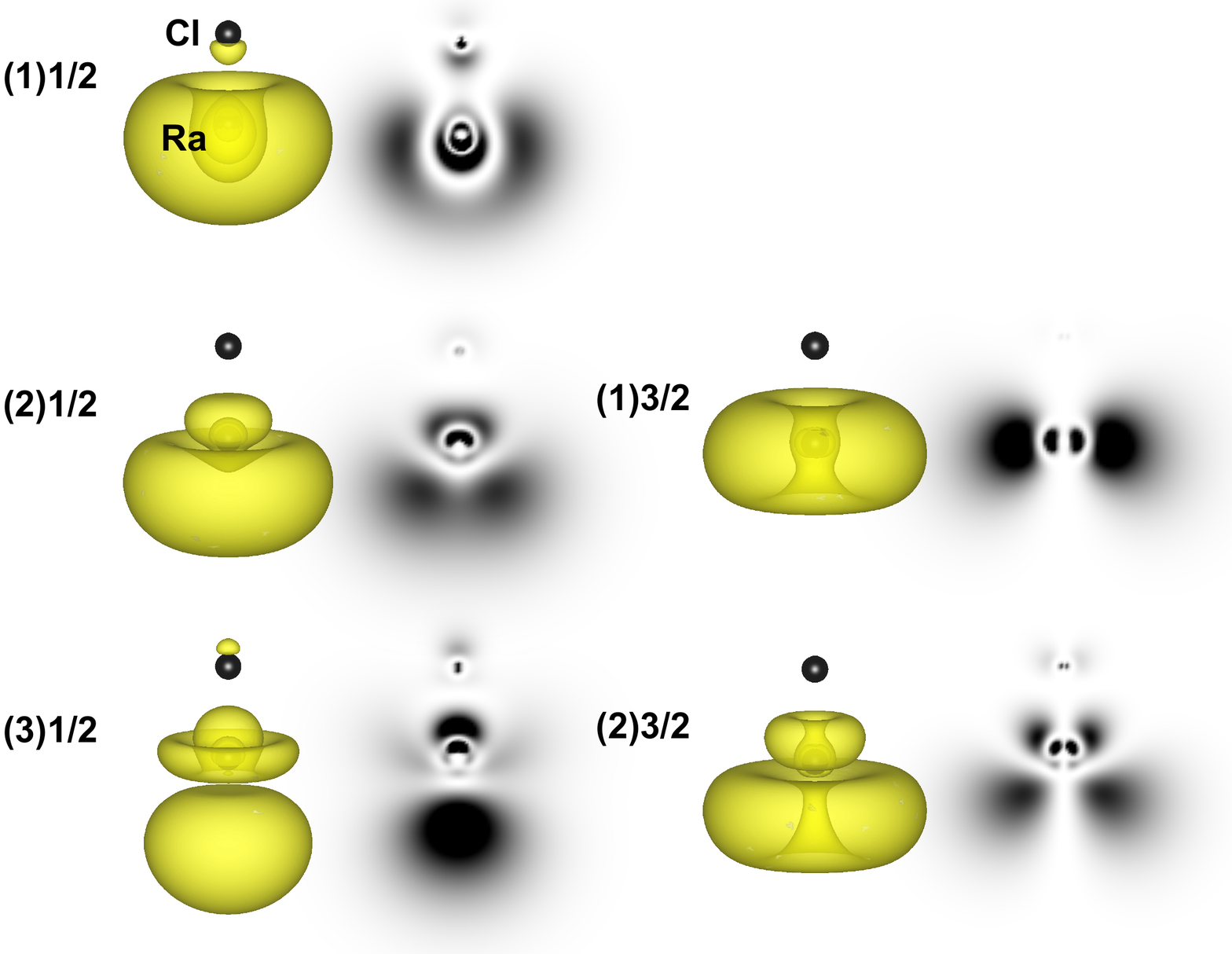}
  \caption{Isodensity surfaces 
 %TI  WHICH  ISODENSITY VALUE WAS SET? 
  and cross sectional view for unpaired-electron pseudonatural spinor densities for low-lying electronic
  states of RaCl. 
 %TI
  Numbers in parenthesis enumerate spinors, whereas the adjacent number corresponds to the $\Omega$ value for each spinor.
 %end TI 
  }
\label{spinors}
\end{figure}

The origin of the parallelism of potential curves %of the ground and low-lying excited states 
can be readily interpreted in the frames of a simple one-electron picture \cite{Isaev:10}.
%TI 
%Note that the model-space projection of FS RCC wavefunctions in the $0h1p$ sector are readily converted to
%a single-determinant form by 
%%an appropriate transformation of active spinors. 
%transformation to natural spinors. The occupation numbers clearly indicate that each electronic 
%state from the model space is essentially a ``one electron over closed shell'' state, with excited electronic states
%being dominated by excitation of the unpaired valence electron to corresponding singly-occupied spinors.
%can be
Note that the model-space projection of any FS RCC wavefunctions in the $0h1p$ sector is immediately and exactly converted to 
a single-determinant form ("one-electron-over-closed-shell") by an appropriate transformation of active spinors. 
Excitations from ground to low-lying excited electronic states are therefore dominated by the promotion of the unpaired valence electron to some higher-lying spinor.
%considered as ``model-space'' approximations for natural transition spinors.
Fig.~\ref{spinors} provides the graphical representation of these ``pseudonatural'' spinors for the ground and low-lying excited states at $R=R_\mathrm{e}(X\,1/2)$. 
%END TI
One immediately notes that all these spinors are localized on the Ra atom and the shape of the corresponding density distributions is typical for non-bonding spinors or orbitals of Class I according to \cite{Isaev:10}, except for the 
energetically  lowest $\Omega=3/2$ spinor, which is rather a Class II spinor. We would like to remind that Class I spinors are spinors which have most 
of their density out of the bonding region, due to e.g. destructive $sp$-hybridisation, and thus electrons occupying such 
spinors do not participate in chemical bonding. Class II spinors are ``atomic-type'' non-bonding spinors, which have a shape of
atomic orbital centered on corresponding atom and cannot be shared between atoms (and thus create a chemical bond), due to 
symmetry reasons (see \cite{Isaev:10, Isaev:16} for details).
     
Calculations of \podd and \ptodd effects in the framework of a quasirelativistic ZORA approach attract considerable interest due to both importance
of these effects for search for new physics (see \cite{Flambaum:85} and \cite{Kozlov:95}) and due to effective test of accuracy of quasirelativistic approaches in
calculations of effects, heavily depending on relativity. In a number of papers \cite{Isaev:12,  Isaev:2014, Gaul:17} it was demonstrated that values 
%RB 
%for ZORA Hamiltonian deviate not more than for 5\% 
from the ZORA approach deviate by less than 5\% 
%(\emph{sure about that number? For heavy ones less than 1\%, especially for closed shell systems. For open-shell we have spin-polarisation})
%TI For W_a calculations we don't have spin-polarisation yet. Difference for Wa in RaF with Borschevsky's result (PRA 2013) is 4.9% 
%(1363Hz(DHF) vs 1300Hz(ZORA))
%
from values obtained with relativistic Dirac-Coulomb Hamiltonian for nuclear spin-dependent \podd 
and scalar electron-nucleus \ptodd interactions. 
Results of our ZORA/GHF and ZORA/GKS calculations of nuclear spin-dependent \podd and scalar electron-nucleus \ptodd parameters are presented in Table \ref{zora}.
%Our absolute value of the $W_{\rm a}$ parameter
%in ZORA/GHF calculations of RaCl is 1.48$\times$10$^3$ Hz, which can be compared with ZORA/GHF value of $W_{\rm a}$ equal to 1.30$\times$10$^3$ Hz for RaF
%\cite{Isaev:16} and 1.38$\times$10$^3$ Hz for RaOH \cite{Isaev:17}. The value of $W_{\rm s}$ is -166 kHz for RaCl, while for RaF it is -150 kHz \cite{Isaev:13} and for 
%RaOH it is -154 kHz \cite{Isaev:17}. 
In both ZORA/GHF and ZORA/GKS calculations, as it is expected
%RB
and as found in previous numerical studies \cite{sunaga:2019},
%RB end
chemical substitution in RaX compounds does not change considerably the magnitudes of 
\podd and \ptodd effects in molecules with non-bonding valence electron. 
%ZORA/GKS values of $W_{\rm a}$ with B3LYP exchange-correlation potential are 1.64$\times$10$^3$ Hz for RaCl and 1.42$\times$10$^3$ Hz for RaF \cite{Isaev:12}.
Our results for RaCl confirm the common trend that accounting for electron correlation and spin-polarisation effects (see also \cite{Isaev:2014} for discussion)
 lead to increase of nuclear spin-dependent \podd effects in open-shell diatomic molecules with $\Sigma_{1/2}$ ground electronic states.
 It is interesting that with \ptodd effects the situation is the opposite -- it was previously observed that accounting for electron correlations decreases the absolute value 
 of $W_{\rm s}$ for about 10\% in RaF (-152 kHz GHF vs. -136 kHz GKS/B3LYP in \cite{Gaul:17b}). The same trend we observe in RaCl, absolute value for $W_{\rm s}$
 drops down to -154 kHz for GKS/B3LYP, which constitutes 7\% decrease from the GHF result. We note that in contrast to our quasi-relativistic complex GHF approach
to those fundamental symmetry violating effects, the paired generalised Dirac--Hartree--Fock (DHF) ansatz used in Ref.~\cite{sunaga:2019} does not capture 
spin-polarisation effects, so that coupled cluster with singles and doubles amplitude (CCSD) gives absolute values for the \ptodd parameters that are
larger than their DHF counterparts.

\begin{table}
\caption{\label{zora} \small Results of ZORA/GHF and ZORA/GKS (with B3LYP exchange-correlation functional) calculations of
$W_{\rm a}$ and $W_{\rm s}$ for the ground electronic $\Sigma_{1/2}$ state of RaCl. } 
 \begin{center}
  \begin{tabular}{lcccccc}
\\
                   &  \multicolumn{3}{c}{$W_{\rm a}$, kHz}  &  \multicolumn{3}{c}{ $W_{\rm s}$, kHz}  \\
  \hline     
         &  RaCl & RaF$^a)$& RaOH$^b)$ & RaCl        & RaF$^c)$    & RaOH$^b)$ \\
ZORA/GHF &  1.48 & 1.30 & 1.38         &$-$166       &$-$152       &$-$154\\
ZORA/GKS &  1.64 & 1.42 & --           &$-$154       &$-$136       & --   \\
% Added by RB
DHF      &   --  & --   & --           &(-)127.9$^d)$&(-)116.9$^d)$& -- \\
CCSD     &   --  & --   & --           &(-)168.1$^d)$&(-)152.5$^d)$& -- \\\hline
  \end{tabular}                                                  
  \end{center}
  $^a)$ data from \cite{Isaev:16}
  $^b)$ data from \cite{Isaev:17}
  $^c)$ data from \cite{Gaul:17b}
% Added by RB
  $^d)$ data from \cite{sunaga:2019}
\end{table}

\section{Conclusion}
The electronic and vibronic structure of diatomic RaCl has been studied with relativistic Fock space coupled cluster methods.
Based on the computed level energies, a reassignment of previously reported levels was proposed, suggesting that the states
formerly assigned as $C,C'$ to a spin-orbit-split ${}^{2}\Pi$ level belong instead to the (2)3/2 level of ${}^{2}\Pi$ character
and the (3)1/2 level of predominantly ${}^2\Sigma$ character. Computed potential energy surfaces for the energetically lowest
electronic states are found to be nearly parallel to each other, suggesting a favourable situation for laser cooling.
In agreement with the experimental report, a very narrow Franck--Condon parabola (nearly diagonal Franck--Condon matrix) is
found and computed branching ratios strongly suggest amenability of RaCl for laser cooling.

As precursor materials for RaCl are more readily available than for RaF, this is excellent news for preparatory off-line studies 
on laser cooling and precision spectroscopy with radium monohalide molecules. Due to medical applications of ${}^{223}$RaCl$_2$ in 
radiotumor therapies, even shorter-lived isotopologues are commercially available, which offers opportunities to study nuclear spin-dependent 
properties and isotope effects with the title molecule.

\section{Acknowledgment}

The authors are grateful to L. Skripnikov and K. Gaul for fruitful discussions. 
The work of T.I. was supported by the grant of Russian Science Foundation N 18-12-00227. R.F.G.R is grateful for financial support 
to DOE Office of Nuclear Physics under grant DE-SC0021179.

\clearpage
%\bibliographystyle{apsrev}
%\bibliography{./az}% Produces the bibliography via BibTeX.

\end{document}